\documentclass[10pt,twocolumn]{article}[1994/06/01]
\usepackage{amsxtra,amssymb,amsthm,amsmath,latexsym}

\theoremstyle{plain}

\def\nd{\noindent}

\def\ve{{\varepsilon}}
\def\R{{\mathbb R}}

\def\barD{{\overline{D}}}
\def\calA{{\mathcal A}}

\def\calN{{\mathcal N}}
\def\calU{{\mathcal U}}

\def\oH{{\overset{\circ}{H}}}
\def\oH1{{\overset{\circ}{H}\kern-.02in{}^1}}

\def\Im{{\hbox{\,Im\,}}}

\def\bee{\begin{equation*}}
\def\eee{\end{equation*}}
\def\be{\begin{equation}}
\def\ee{\end{equation}}

\begin{document}

\pagestyle{empty}
%\begin{titlepage}
\title{
\textbf{CREATING WAVE-FOCUSING MATERIALS}
}
\author{\textbf{A.G. Ramm}\\
 Department of Mathematics \\
 Kansas State University, \\
 Manhattan, KS 66506-2602, USA\\
 ramm@math.ksu.edu
\\ fax 785-532-0546, tel. 785-532-0580}
%http://www.math.ksu.edu/\,$\widetilde{\ }$\,ramm}

\date{}
\maketitle\thispagestyle{empty}
\par\par

\nd\textbf{ABSTRACT}

%\vskip-.02in

Basic ideas for creating wave-focusing materials by injecting small
particles in a given material are described.

The number of small particles to be injected around any point is
calculated. Inverse scattering problem with fixed wavenumber and fixed
incident direction of the plane acoustic wave is formulated and solved.
\par \par

%\end{titlepage}
\vskip.1in
\nd\textbf{1.~INTRODUCTION}

This paper contains the results presented at the author's plenary talk at the 
IPDO-2007 symposium
on inverse problems, design and optimization. A method for creating
materials with a desired refraction coefficient is given. This coefficient,
in particular, may be chosen so that the new material has a desired 
wave-focusing property (see also [1]-[12]).

We want to investigate the following problem. Let $D$ be a bounded domain
filled with a material whose properties are known, for example, a
homogeneous material with known speed propagation of sound waves. Can one
inject into $D$ small particles in such a way that the resulting new
material would have some desirable wave-focusing properties? For example,
is it possible to create in this way a material that scatters an incident
plane wave in a desired solid angle?

There is a large engineering and physical literature on creating ``smart"
materials. Photonic crystals, quantum dots, coating, are some key words.
However, there seems to be no prior work which deals with the question
posed above.

In this paper this question is studied rigorously: the ``smallness" of
the particles is specified, the number of these particles around the point
$x$ is specified, the role of the shapes of these particles is explained,
and the notion of wave-focusing is made precise.

The basic results of this paper are:

 1) It is proved that the injection of a suitable number of small,
acoustically soft particles in a given bounded region, filled by some
material with known properties, allows one to create a new material such
that its scattering amplitude is arbitrarily close to a given scattering
amplitude. \\
 2) A method is given to calculate the density $N(x)$ of small
particles, to be injected in a unit volume around a point $x\in D$, in
order that the new material has the scattering amplitude close to the
desired scattering amplitude. \\
 3) For the first time the problem of finding a compactly supported
potential $q(x)$ which generates the scattering amplitude
$A(\beta):=A(\beta,\alpha,k),$ approximating an arbitrary fixed given 
function $f(\beta)\in
L^2(S^2)$ with any desired accuracy is formulated and solved. Here the 
wavenumber $k>0$ and the
incident direction $\alpha\in S^2$ are fixed, $S^2$ is the unit sphere in
$R^3$.

In Section 2 detailed statements of the problems are given. In Section 3
the inverse scattering problem with fixed wavenumber $k>0$ and fixed
incident direction $\alpha\in S^2$ is discussed.

In Section 4 the ill-posedness of the above problems is discussed.

For engineers the paper gives a "recipe" for creating materials
with a desired refraction coefficient. No such receipes were 
given earlier, to the author's knowledge, although there were 
many papers (see book [13] and references therein) in which an effect of 
embedding small spheres or ellipsoids into
a homogeneous material on the effective dielectric and magnetic
properties of the material was discussed, and the new material
was a homogenized material. Homogenization in the literature
was studied mostly for elliptic positive-definite  operators,
while we study wave propagation and the 
corresponding operator is not positive-definite. 

Another principal difference between this paper and the earlier
published results consists in the statement of the problem:
we are not considering the distribution of small particles,
embedded into the material, as uniform or random, but 
solve a design problem of creating materials with a desired refraction 
coefficient. 

Moreover, we give (see also [10]-[12]) a precise recipe for creating
such material for practically arbitrary refraction coefficient.

The two technological (engineering) problems to be solved    
for this recipe to be practically implemented, can be 
formulated precisely as well: 

a) How does one practically embed in a given
material many small particles given the number of the particles per unit 
volume around every point of the original material?

b) How does one prepare practically small particles with
a desired boundary impedance?

In this paper we consider small particles with the Dirichlet 
boundary condition, corresponding to acoustically soft particles.
Problem b) 
will arise if one considers the small particles on the 
boundary of which an impedance boundary condition is imposed.
Varying the boundary impedance as a function of positions of 
particles, one can create refraction coefficients with the 
desired absorption properties. This was discussed in more detail in 
[12].

\par\par

\vskip.1in
\nd{\textbf{2.~STATEMENT OF THE PROBLEM AND SOME RESULTS}

Consider a bounded domain $D\subset\R^3$ with a smooth boundary $S$. The
scattering of a plane wave on this domain is described by the equations
\be\label{e1} [\nabla^2+k^2n^2_0(x)]u=0\hbox{\ in\ } D, \ k=const>0,\ee
\be\label{e2}
	\begin{aligned} 
	u&=e^{ik\alpha\cdot x}+A_0(\beta,\alpha,k)
	\,\frac{e^{ikr}}{r}+o\left(\frac{1}{r}\right), \\
	&r:=|x|\to\infty, \quad \beta:=\frac{x}{r}.\end{aligned}\ee 
 Here
$\alpha\in S^2$ is given, $A_0(\beta,\alpha,k)$ is the scattering
amplitude, $k>0$ is fixed throughout the paper, $n^2_0(x)>0$ is a given
function, the refraction coefficient, $n^2_0(x)=1$ in $D':=R^3\backslash 
D$,
$n^2_0(x)$ is piecewise-continuous. The function $n^2_0(x)$ describes the 
material
properties of the region $D$. Problem (1)--(2) has a unique solution $u\in
H^2_{loc}(R^3)$, where $H^2_{loc}(R^3)$ is the Sobolev space. Equation (1) can 
be written as a Schr\"odinger equation
$$[\nabla^2+k^2-q_0(x)]u=0,\quad q_0(x):=k^2[1-n^2_0(x)],$$ $q_0=0$ in 
$D'$.
Suppose that $M$ small acoustically soft particles (bodies) $D_m$ are
injected into domain $D$. Smallness means that $ka << 1$, where $a$ is the
characteristic size of the small bodies. One may define $a:=\frac 1 
2\max_{1\leq
m\leq M} diam D_m$. Assume that the boundaries $S_m$ of $D_m$ are uniformly
Lipschitz, i.e., the Lipschitz constant does not depend on $m$.
Acoustically soft means that $u|_{S_m}=0,$ where $u$ can be interpreted as
acoustic pressure.

The scattering problem in the new region can be formulated as follows:
\be\label{e3}
  [\nabla^2+k^2 n^2_0(x)]\calU=0\hbox{\ in\ }R^3\setminus 
\bigcup^M_{m=1}D_m, \ee
\be\label{e4} \calU|_{S_m}=0, \quad 1\leq m\leq M, \ee
\be\label{e5}
\begin{aligned}
  \calU =u+& \calA_M(\beta,\alpha,k)
	\frac{e^{ikr}}{r}+o\left(\frac{1}{r}\right), \\
  &r\to\infty,\ \frac{x}{r}=\beta, \end{aligned}\ee
 where $u$ solves (1)-(2). One can replace (5) by the following:
\be\label{e6}
\begin{aligned}
  \calU=e^{ik\alpha\cdot x}
  &+A(\beta,\alpha,k)\frac{e^{ikr}}{r} + o\left(\frac{1}{r}\right) \\
  & r\to\infty, \quad \frac{x}{r}=\beta.\end{aligned}\ee
From (2), (5) and (6) one gets $A=A_0+\calA_M$.

Since $n^2_0(x)$ is known, $A_0$ is known.

Problem (3)-(5) has a unique solution.

{\bf Problem 1} is to show that one can distribute sufficiently large
number $M$ of small particles in $D$ in such a way that
$A(\beta, \alpha):=A(\beta,\alpha,k)$, $k>0$ is fixed, will approximate
 in $L^2(S^2)$ an
arbitrary fixed scattering amplitude  $f(\beta, \alpha)$ with any 
desired accuracy.  
We consider also the following problem.\\ 
{\bf Problem 2:} Can one distribute small particles in
$D$ so that the resulting new material would have the scattering amplitude
$A(\beta)$  ($k>0$ and $\alpha\in S^2$ are both now fixed), which
approximates in $L^2(S^2)$ an arbitrary given function $f(\beta)\in
L^2(S^2)$ with any desired accuracy?

If $A(\beta,\alpha):=A(\beta,\alpha,k)$ ($k>0$ is fixed) is a scattering
amplitude, known for all $\beta,\alpha \in S^2$, then one can find the
unique, corresponding to $A(\beta,\alpha),$ potential $q(x)$ by the Ramm's
method \cite{R425}, \cite{R470}. This method gives a stable approximation of 
$q$ even
in the case when noisy data $A_\delta(\beta,\alpha)$ are given,
$\sup_{\beta,\alpha\in S^2}
|A(\beta,\alpha)-A_\delta(\beta,\alpha|<\delta$.

If $q(x)$ is found from $A(\beta,\alpha)$, then we define $p(x):=q-q_0(x)$
and prove that $p(x)=N(x)C_0$, where $N(x)$ is the number of small
particles per unit volume around a point $x$, i.~e.~, the spatial density
of the number of the particles, and $C_0$ is the electrical capacitance of
a small conductor of the same shape as the particle. Here we assume that
all the small particles are identical, but this assumption can be
dropped (see  \cite{R492},  \cite{R515} ).

{\bf Let us summarize:} If one injects small particles with the spatial
density $N(x)=\frac{p(x)}{C_0},$ where $p(x):=q(x)-q_0(x)$, then the
resulting new material will have practically the desired scattering amplitude
$A(\beta,\alpha)$, corresponding to a potential $q(x)$.

This gives a solution to {\bf Problem 1}. Note that the scattering
amplitude $A(\beta,\alpha)$, $\forall \beta,\alpha\in S^2$, corresponding
to a real-valued potential $q\in L^2(D)$, determines $q$ uniquely. Since
$q_0(x)$ is known, the function 
$$N(x)=\frac{q(x)-q_0(x)}{C_0}$$ gives an
approximate solution to {\bf Problem 1}. This solution is not exact
because we did not pass to the limit 
$$M\to\infty, \quad ka\to 0,\quad \frac{a}{d}\to 0,$$ 
but took just sufficiently small identical particles
with $C_0$ being electrical capacitance of the perfect conductor of the
same shape as a single particle.

The number of small particles per unit volume is $O(\frac{1}{d^3})$. Their
volume per unit volume of the original medium
is $O(\frac{a^3}{d^3})$. This quantity tends to
zero as $\frac{a}{d}\to 0$. The capacitance per unit volume is
$O(\frac{a}{d^3})$. Thus, the limit of the ratio $\frac a {d^3}$ is 
finite
and non-zero, while the limit of the relative volume of the injected
small particles is zero because $\frac {a^3}{d^3}$ tends to zero.
\par\par

\vskip.1in
\nd\textbf{3.~INVERSE SCATTERING WITH FIXED $k$ AND $\alpha$}

Our solution to {\bf Problem 2} is based on the idea used in Section 2 in 
solving {\bf Problem 1}. 
Given $f(\beta)$, we find $q(x)\in L^2(D)$, such that the corresponding to $q$ 
scattering amplitude $A_q(\beta)$ ($k>0$ and $\alpha\in S^2$ are fixed)
approximates $f(\beta)$ with a desired accuracy: 
$\| f(\beta)-A_q(\beta)\|_{L^2(D^2)}<\ve$, 
where $\ve>0$ is an a priori given small number.
If such a $q$ is found, then $N(x)=\frac{q(x)-q_0(x)}{C_0}$ as in Section 2.
The principally novel problem is finding $q$ from $f(\beta)$ and $\ve$.
This problem has many solutions, as we prove. However, a priori it is not at 
all clear 
if this problem has a solution. Let us outline our solution to this problem. 
First, recall the well-known exact formula for the scattering amplitude: 
$$A_q(\beta)=-\frac{1}{4\pi}\int_D e^{-ik\beta\cdot x} q(x)u(x)dx, $$
where $u(x)$ is the scattering solution,
$$[\nabla^2+k^2-q(x)]u=0 \hbox {\,\, in \,\,} R^3,$$ 
$$u=u_0+A_q(\beta)\frac{e^{ikr}}{r}+o(\frac{1}{r}), \quad r\to\infty, $$ 
$\beta=\frac{x}{r}$, $u_0:=e^{ik\alpha\cdot x}$, $\alpha\in S^2$ and
$k>0$ are fixed, $u(x)=u(x,\alpha,k)$.
Denote $h(x)=q(x)u(x)$. 
Then 
$$A_q=-\frac{1}{4\pi}\int_D e^{-ik\beta\cdot x} h(x)dx,$$
where $A_q$ is the scattering amplitude, corresponding to the 
potential $q$.
Given $f(\beta)$ and $\ve>0$, however small, one can find (many) $h$ such 
that
\be\label{e7}
  \|f(\beta)+\frac{1}{4\pi}\int_D e^{-ik\beta\cdot x}
  h(x)dx\|_{L^2(S^2)}<\ve. \ee
 The function $h$ can be found, for example, as a linear combination 
$h_n=\sum^n_{j=1}c_j\varphi_j(x)$, where$\{\varphi_j\}$ is a basis of 
$L^2(D)$. If $n$ is sufficiently large and $c_j$ are found from the 
minimization problem
$$\|f+ \sum^n_{j=1}\frac {c_j}{4\pi} \int_D e^{-ik\beta\cdot 
x} 
\varphi_j(x)dx\|_{L^2(S^2)}=\\
min,$$ 
then (7) holds.
In \cite{R515} another, analytical, solution to (7) is given.

If $h=h_\ve(x)$ is found, then $q(x)$ can be found from the nonlinear 
equation $h=qu$. This equation for $q$ is nonlinear because the scattering 
solution $u=u(x;q)$  depends nonlinearly on $q$. One has
\be\label{e8}
  u(x)=u_0(x)-\int_D g(x,y)q(y)u(y)dy, 
\ee
where $u_0=e^{ik\alpha \cdot x}$ and $ g:=\frac{e^{ik|x-y|}}{4\pi|x-y|}$.
 Let
\be\label{e9}
  q(x):=\frac{h(x)}{u_0(x)-\int_D g(x,y)h(y)dy}. \ee
If the right side of (9) is an $L^2(D)$ function, then (9) solves our 
inverse scattering problem. Indeed, define
$$u(x):=u_0(x)-\int_D g(x,y)h(y)dy.$$ This $u$ solves (8) with $q$ defined 
in (9). The scattering amplitude 
$$A_q(\beta)=-\frac{1}{4\pi} \int_D e^{-ik\beta\cdot x} h(x)dx.$$
By (7) one has
$$\|f(\beta)-A_q(\beta)\|_{L^2(S^2)}<\ve.$$
 So, {\bf Problem 2} is solved 
if (9) defines an $L^2(D)$ function.
This, for example, is the case if 
$$\inf_{x\in D}|u_0(x)-\int_D g(x,y)h(y)dy|>0.$$
If formula (9) defines a non-integrable function due to possible zero 
sets of the function $$\psi(x):=u_0-\int_D g(x,y)h(y)dy, $$
then, as we prove, a suitable small perturbation $h_\delta$ of $h$ in 
$L^2(D)\hbox{-norm}$ will lead to a function 
$$q_\delta:=\frac{h_\delta}{\psi_\delta}\in L^2(D),$$
where
$$\psi_\delta=u_0-\int_D g(x,y)h_\delta (y)dy,$$
and 
$\|h-h_\delta\|_{L^2(D)}<\delta$.
Since the function 
$-\frac{1}{4\pi} \int_D e^{-ik\beta\cdot x}h_\delta(x)dx$ 
differs a little from the function 
$-\frac{1}{4\pi} \int_D e^{-ik\beta\cdot x}h(x)dx$,
condition (7) is satisfied if $h$ is replaced by $h_\delta$ 
and $\ve$ by, for example,
$2\ve$. Therefore 
$$N_\delta(x):=\frac{q_\delta(x)-q_0(x)}{C_0}$$ 
solves {\bf Problem 2} in the case when (9) is not an 
$L^2(D)\hbox{-function}$.

Let us explain how to choose $h_\delta$. First, without loss of 
generality 
one can assume $h$ to be analytic in $\barD=D\bigcup S$, because analytic 
functions (even polynomials if $D$ is bounded) are dense in $L^2(D)$. If 
$h$ is analytic, so is $\psi(x)$ in $D$. Therefore the null set of $\psi$,
$$\calN:=\{x:\psi(x)=0,\,\,\, x\in D\},$$ is generically a line defined by 
two equations 
$$\psi_1:=\hbox{Re\,}\psi=0, \quad \psi_2:=\Im\psi=0.$$
Let 
$$\calN_\delta:=\{x:|\psi|<\delta,\, x\in D\},$$
and $D_\delta:=D\backslash \calN_\delta$.
Generically $|\nabla\psi|\geq c>0$ on $\calN$ and, by continuity,
this inequality holds in 
$\calN_\delta$ (possibly with a different $c>0$). Small perturbation of 
$h$ leads to these generic assumptions.

Define
\be\label{e10}
\begin{aligned}
  h_\delta=&\left\{\begin{array}{l}
     h\hbox{\ in\ }D_\delta,\\0\hbox{\ in\ }\calN_\delta,
  \end{array}\right.\\
  q_\delta:=&\left\{\begin{array}{ll}
    \frac{h_\delta(x)}{u_0-\int_D g(x,y)h_\delta(y)dy}
     & \hbox{\ in\ }D_\delta,\\
   0 & \hbox{\ in\ }\calN_\delta.\end{array}\right.
\end{aligned}\ee
 Then $q_\delta\in L^2(D)$ and 
$$N_\delta(x):=\frac{q_\delta(x)-q_0(x)}{C_0}$$ 
solves {\bf Problem 2}.

Let us check that $q_\delta\in L^2(D)$. We prove more: 
$q_\delta\in L^\infty(D)$. It is sufficient to check that
\be\label{e11}
  \begin{aligned}
  \inf_{x\in D_\delta} &|\psi_\delta(x)|=\\
  \inf_{x\in D_\delta} &|u_0(x)-\int_{D_\delta} g(x,y)h(y)dy|
  \geq c \delta>0,\end{aligned}\ee
 because $q_\delta=0$ in $\calN_\delta$ by the definition.

Choose the origin on $\calN$ and make a change of variables
\be\label{e12}
  s_1=\psi_1(x),\, s_2=\psi_2(x),\, s_3=x_3. \ee
 The Jacobian of this transformation of variables is non-singular because 
$\nabla\psi_1$ and $\nabla\psi_2$ 
are linearly independent on $\calN$ and in $\calN_\delta$. We have 
$$\max_{x\in \calN_\delta} (|J|+|J^{-1}|)\leq c,$$ $c>0$ stands for a 
generic constant independent of $\delta$. Let us check that
$$|\psi_\delta(x)|\geq c\delta \hbox{\quad in \quad} D_\delta.$$ 
We have $$|\psi_\delta(x)|\geq |\psi(x)|-I(\delta),$$
where $$I(\delta)=\int_{\calN_\delta} |g(x,y)h(y)|dy.$$ 
If $x\in D_\delta$, then $|\psi|\geq\delta$, and
$$|\psi_\delta(x)|\geq\delta-I(\delta).$$ 
Moreover, $\max_{y\in D}|h|\leq 
M$ 
and, using the new variables (12), one gets 
$$I(\delta)\leq 
c\int_{\rho\leq\delta}d\rho\,\rho\int^1_0\frac{ds_3}{\sqrt{s^2_3+\rho^2}},$$
because the region $\calN_\delta$ can be described by the inequalities 
$$\rho^2=s^2_1+s^2_2\leq\delta^2, \quad 0\leq s_3\leq 1,$$
and we have used the estimate $|J^{-1}|\leq c$ in $\calN_\delta$.
Integral 
$$I(\delta)=O(\delta^2|\ln \delta|) \hbox {\quad as \quad}\delta\to 
0.$$
Thus $$|\psi_\delta|\geq\delta-O(\delta^2|\ln \delta|)\geq c\delta$$ 
with some constant $c\in (0,1)$.
This justifies our method for solving {\bf Problem 2} in the case when 
formula (9) does not yield $q\in L^2(D)$.
\par\par

\vskip.1in
\nd\textbf{4.~ILL-POSEDNESS OF PROBLEMS 1 AND 2}

Both {\bf Problems} 1 and 2 are ill-posed. Since the ill-posedness of 
{\bf Problem 1}
has been discussed in great detail in \cite{R425},  
\cite{R470}, we discuss
only the ill-posedness of {\bf Problem 2}.

In {\bf Problem 2} one has to find $h$, given $f$ and $\ve>0$, so that (7) 
holds. This is an ill-posed problem, similar to solving the first kind 
integral equation
$$Bh:=-\frac{1}{4\pi} \int_D e^{-ik\beta\cdot x} h(x)dx=f(\beta)$$ for 
$h$.
If this equation is solvable for a given $f$, it may be not solvable if 
$f$ is replaced by a slightly perturbed function $f_\delta$. If $\ve>0$ in 
(7) is small and $f$ is not in the range of $B$, then $\|h\|_{L^2(D)}$ is 
large. This leads to large maximal values of the corresponding $q$.
Therefore any numerical method for solving {\bf Problem 2} has to use a 
regularization procedure. In \cite{R521} one can find some numerical 
results 
related to {\bf Problem 2} and a description of the regularization 
procedure 
which was used.

\vspace{-.075in}
\renewcommand{\refname}{\nd\normalsize{

\end{document}